# Deep learning guided Android malware and anomaly detection

Nikola Milosevic[1], Junfan Huang[2]

[1] The Department of Computer Science, School of Engineering, University of Manchester, Manchester, UK

[2] The University of Bristol, Bristol, UK

## Abstract
In the past decade, the cybercrime related to mobile devices has increased. Mobile devices, especially the ones running on Android operating system are particularly interesting to malware creators, as the users often keep the biggest amount of personal information on their mobile devices, such as their contacts, social media profiles, emails, and bank accounts. Both dynamic and static malware analysis is necessary to prevent and detect malware, as both techniques have their benefits and shortcomings. In this paper, we propose a deep learning technique that relies on LSTM and encoder-decoder neural network architectures for dynamic malware analysis based on CPU, memory and battery usage. The proposed system is able to detect and notify users about anomalies in system that is likely consequence of malware behaviour. The method was implemented as a part of OWASP Seraphimdroid's anti-malware mechanism and notifies users about anomalies on their devices. The method proved to perform with an F1-score of 79.2%.

## Introduction
Mobile devices are a popular target for cyber-crime, since they store a large amount of personal information, including contacts, accounts on social media and banking details. Some of this information is stored in installed applications, while others may be stored in documents on the device. In either case, if malware is installed, the attacker may gain access to this information and exploit them. According to GSMA, there are over 5.1 billion subscribers to mobile services (GSMA, 2019). This means that 67% of the global population have a mobile device and the number will be growing in the years to come. In 2019, there have been 3.4 billion people connected to 4G networks around the globe. By the year 2025, expectations are that around 25 billion IoT devices will be connected to the network. The most represented operating system for mobile devices and IoT is Android, powering not only mobile phones, but also smart TVs and other devices.

The Android operating system has the biggest market shares among mobile devices and it is dominating the mobile operating systems market since 2010 (Alfred & Kaijage, 2019). At the moment, about 86.8% of mobile devices in the world are working on the Android operating system (Qiu, et al., 2019). This makes Android the most attractive platform for malware developers. On the other hand, mobile application stores are fighting against malware applications being published there with variable success. However, lately, malware developers came up with the approaches to spread using SMS and other methods, and therefore avoiding application stores completely

(McAffee, 2019).  The fact that users could install applications from various non-official stores, that may not employ security and anti-malware testing in the process of approving and publishing applications puts mobile users in a dangerous position. The spread of malware that is not relying on any store further moves the main venue for fighting malware to users' mobile devices.

According to McAfee Mobile Threat Report that was published in 2019, an average person has installed between 60 and 90 applications on their phone. Among these applications may be applications that have a malicious purpose. According to the same report, the emergence of hidden applications, that installs on the phone and do not present themselves to the users may present additional threat that user can overcome only with a proper methodology for malware detection.

In our previous paper (Milosevic, Dehghantanha, & Choo, 2017), we have described our approach taken by the OWASP Seraphimdroid app for static analysis. In this paper, we present a methodology for dynamic malware analysis that was developed as a part of the OWASP Seraphindroid[1] project. The presented work has been developed under the OWASP Seraphimdroid supervision over Google Summer of Code 2019.

# Background

Existing approaches for malware detection and analysis can be categorized into dynamic and static malware analysis. In static analysis, one inspects the source code and binaries in order to find suspicious patterns. Dynamic analysis (behavioural-based analysis) involves the execution of the malware in an isolated environment while monitoring and tracing its behaviour (Schmidt, Clausen, Camtepe, & Albayrak, 2009). In this paper, we will focus on reviewing machine learning aided approaches to malware detection using dynamic analysis.

Approaches to dynamic mobile malware detection that were proposed in the past relied on detecting anomalies in battery consumption, which could be caused by malware activities (T.K. Buennemeyer, 2008; Kim, Shin, & P., 2011). API calls, I/O requests, resource locks and battery consumption are valuable for dynamic malware detection (Blaasing, Batyuk, Schmidt, Camtepe, & S., 2010). TaintDroid is a malware detection system that monitors the application's behaviour and detects anomalies (Enck, et al., 2014). M0Droid is analysing system calls of Android applications on the server and creating signatures. Signatures are later sent to the devices so they can warn the users about threats (Damshenas et al. 2015).

Machine learning aided approaches for dynamic Android malware analysis are quite rare. Chen et al. used a combination of static (permissions, API calls) and dynamic (network activity, file system access, interaction with the operating system) features as input to traditional machine learning algorithms (SVM, decision trees, Naïve Bayes) in order to classify malware. They used DroidBox application in order to obtain features from dynamic analysis (Chen, Xue, Tang, Xu, & & Zhu, 2016). The obtained accuracy was about 93%. DroidBox is supported for Android API version 16 (Android 4.1) and due to the changes in some of the newer API versions, it won't be able to extract much information from devices running on them. DroidDetector utilized as well the combination of static features and dynamic features obtained from DroidBox as input do Deep Belief Networks and improved the results of malware classification to about 96% (Yuan, Lu, & Xue, 2016). Mobile-sandbox utilizes the SVM machine learning algorithm to guide code execution and gather information about native code execution (Spreitzenbarth, Schreck, Echtler, Arp, & Hoffmann, 2015). Features obtained about API and system calls were used as input to dynamic analysis-based malware classifier (Afonso, Amorim, Grégio, Junquera, & Geus, 2015).

---

[1] https://www.owasp.org/index.php/OWASP_SeraphimDroid_Project

The effect of emulator-based and phone-based malware analysis was analysed and the results showed that about 24% more applications were successfully analysed on the phone. This is due to the restriction that emulator and virtualization impose (Alzaylaee, Yerima, & & Sezer, 2017).

# Method

During the Google Summer of Code 2019, we have developed an approach that is based on detecting anomalies on infected Android systems by monitoring network traffic, battery usage, CPU and memory consumption. The approach is based on encoder-decoder and LSTM neural networks that had significant successes in areas of natural language processing, especially machine translation.

## Data collection

The idea of the project was to collect some of the system information and detect anomalies in case of abnormal usage of the system. We have merged code from another open source applications called AnotherMonitor[2] and NetworkMonitor[3]. Another Monitor helped us collect CPU usage, memory usage and cached information, while Network Monitor provided functions to obtain system battery status and other network information.

For collecting data sets, we have run 40 applications on the emulator for about an hour and collected recorded information. Half of these applications were malicious applications from M0Droid dataset (Damshenas, Dehghantanha, Choo, & Mahmud, 2015), while the other half were benign applications, mainly commonly used applications such as Youtube, Gmail, Chrome, FaceApp, etc.

Compared with benign, our hypothesis is that malware's behaviour would cause the abnormal device usage, for example, CPU used rapidly, battery ran out quicker. Since OWASP Seraphimdroid is able to record these information, it is possible to detect malicious behaviours based on previous behaviours.

## Machine learning models

In our research, we have experimented with an ensemble of two kinds of deep learning architectures. The first is based on encoder-decoder architecture (Vincent, Larochelle, Lajoie, Bengio, & Manzagol, 2010; Cho, Van Merriënboer, Bahdanau, & Bengio, 2010), while the other is based on Long-Short term memory (LSTM) recurrent neural networks (Gers, Schmidhuber, & Cummins, 1999).

Encoder-decoder is a special neural network that encodes the real data into a small dimension, and then decodes it back (see Figure 1). After this operation, it could usually recover well normal data. However, if it is a malicious behaviour, it is possible to obtain a large distance between model prediction and real data. Encoder-decoder architecture could recover the normal data well because it follows some of the distributions, while abnormal data would deviate from that distribution.

---

[2] https://github.com/AntonioRedondo/AnotherMonitor

[3] https://github.com/caarmen/network-monitor

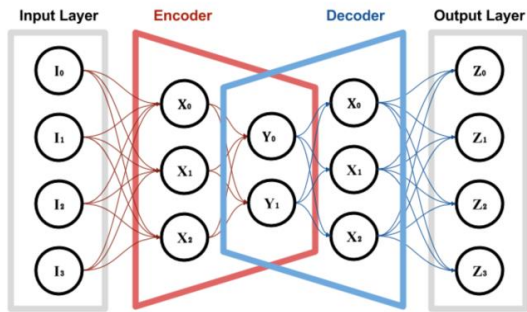

Figure 1: General architecture of encoder-decoder architecture

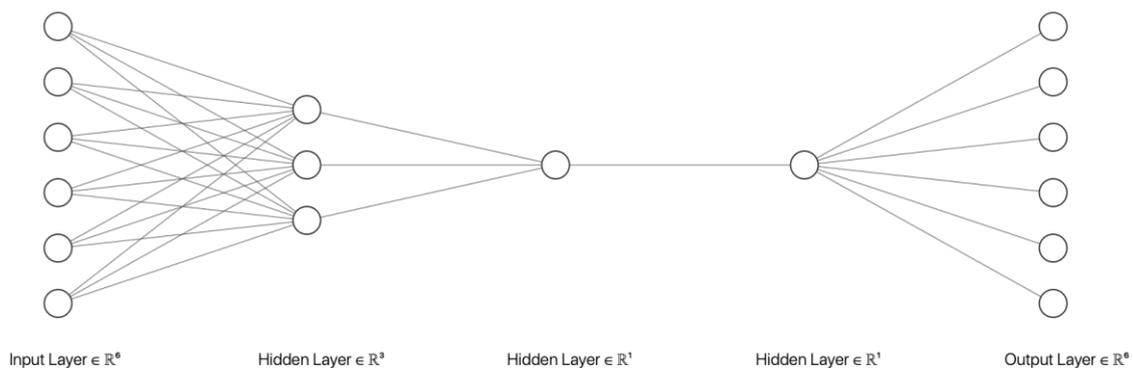

Figure 2: The OWASP Seraphimdroid implementation of encoder-decoder network

AnotherMonitor provides us CPU usage and Memory usage information which are used as inputs to our encoder. In total 6 values are passed to our encoder (i.e. CPU usage, Seraphimdroid CPU usage, free memory, etc.). The implemented architecture is presented in Figure 2. Both inputs and outputs are represented recorded values. As it has been shown in the picture above, the naive autoencoder model contains 2 encoder layers and 2 decoder layers. Once data are collected with these 6 features, they would be standardised by their means and standard deviations. The standardised data would be feed into the neural network above. For the encoding process, the data will be encoded from 6 dimensions into 1 dimension after going through the first two encoders.. For the decoding process, the aim is to recover the encoded data. After passing through two decoders, the zipped data would be recovered back to 6 dimensions. The distance between the outputs (prediction) and the inputs (data) could be used for anomaly detection. The distance metric used in our implementation was Euclidian distance. Usually, the distance would be relatively small, however, once anomaly happens, the prediction would be unstable which would cause a bigger distance between inputs and outputs. Empirically, we have set a threshold to 100 in our case.

The other method we have developed is by using LSTM recurrent neural networks. The input to the LSTM network is M (in our implementation case – 20) previous measurements (from 20 past seconds). The network is trying to predict $21^{st}$ (M+1th) data points. By calculating the distance between prediction and real data and comparing with threshold, it could detect the anomaly.

The anomaly is detected in case that the distances of both models is larger than 100 (encoder_dis>100 and LSTM_dis>100). We have created also a model that sets threshold of output distance on 50 for either of the model (encoder_dis>50 or LSTM_dis>50).

The model was trained on system data for 15 minutes while some applications were run. The training was trying to learn how to predict feature vector (next one in LSTM and the current in encoder-decoder). This model was then used to calculate feature vector that was compared with the real values. The distance between these vectors was compared against threshold and in case the distance is larger than 100 for both models the notification about the anomaly in the system was raised.

# Results

The encoder-decoder architecture was evaluated on 20 applications, out of which 10 were benign and other 10 were malicious. We have run these applications for 10 minutes each and counted the number of anomalies that were reported by the system. As the initial threshold was reporting too many anomalies for non-malicious applications, we have empirically adjusted the threshold, until the balance between anomalies reported by the system did not balance and minimized false positive reports.

The results of 20 application runs are presented in the table below (Table 1).

Table 1: The results of encoder-decoder architecture evaluation on 20 applications. The presented model uses a combination of models, where both models have distance between the predicted feature vector and the real feature vector higher than 100 (encoder_dis>100 and LSTM>100)

| Malware name | Warnings | Benign name | Warnings |
| --- | --- | --- | --- |
| Ebay SMS | 2 | Chrome | 0 |
| Walpaper SMS | 3 | YouTube | 0 |
| Lock your phone | 0 | Temple run | 0 |
| 01B1…apk | 5 | Video Cutter | 0 |
| Android Security Suite premium | 0 | All video downloader | |
| Install | 0 | Video joiner | 0 |
| A79..apk | 2 | Power director | 0 |
| Afa..apk | 1 | Gmail | 0 |
| Blackmart Alpha | 0 | FaceApp | 0 |
| Calendar | 0 | Mr Bullet Spy Puzzles | 0 |

As it can be seen from the table, the method did not produce any false positive, and managed to alarm about 6 out of 10 malicious applications in the first 10 minutes of their execution. Four malicious applications were not detected using this method, and this may be due to the delayed malicious action. In some cases malware will wait and not perform any malicious action for long periods of times.

Table 2: Performance of the model in terms of precision, recall and F1-score.

| | Precision | Recall | F1-score |
| --- | --- | --- | --- |
| **Benign** | 71.4% | 100% | 83.3% |
| **Malicious** | 100% | 60% | 75 % |
| **Overall (macro-average)** | 85.7% | 80 | 79.2% |

In Table 2 is presented model performance in terms of precision, recall and F1-score. As it can be seen from the table, the model of totally unsupervised model is able to detect whether there are malicious behaviour or not in about 80% of cases. The performance might have been higher, in case the test was run for longer time, as some of the malwares were not reported as they may not have started their malicious sequence.

## Discussion

In this research, we presented an ensemble deep learning method based on encoder-decoder architecture and LSTM networks. The method is fully unsupervised and completely relies on the system and data it is run on. Based on this, the system learns the usual CPU, memory, battery and network usage patterns. In case of big deviations from these patterns the anomaly notification is raised and user can react to it.

The presented system performed with 80% F1-score. We would consider that the presented performance is quite good for behaviour only based system. We need to note that our test was based on 10 minute run of applications and it is possible that during these 10 minutes malicious sequence was not triggered. On the other hand OWASP Seraphimdroid has implemented static malware analysis method that examines apps based on their permissions. In the combination with static component of the OWASP Seraphimdroid that already exists the proposed system adds value, as OWASP Seraphimdroid is becoming all rounded security application equipped with both static and dynamic analysis tools on device.

The method can be potentially further improved by pre-training on larger datasets and then fine-tuning on device. Also, combination of supervised and unsupervised methods can further improve the performance. We are also planning to do more extensive and longer test runs.